%% file: ms.tex
\documentclass[12pt,preprint]{aastex}
\usepackage{amsmath}

\shorttitle{QSO Variability Selection}
\shortauthors{Butler et al.}

\def\gtrsim{\mathrel{\hbox{\rlap{\hbox{\lower4pt\hbox{$\sim$}}}\hbox{$>$}}}}
\def\lessim{\mathrel{\hbox{\rlap{\hbox{\lower4pt\hbox{$\sim$}}}\hbox{$<$}}}}

\slugcomment{Accepted to AJ}

\begin{document}

\setlength{\pdfpageheight}{\paperheight}
\setlength{\pdfpagewidth}{\paperwidth}

\title{Optimal Time-Series Selection of Quasars}

\author{Nathaniel R. Butler\altaffilmark{1}
\& Joshua S. Bloom\altaffilmark{2,3}}
\altaffiltext{1}{Einstein Fellow, Astronomy Department,
University of California, Berkeley, CA, 94720-7450, USA}
\altaffiltext{2}{Astronomy Department,
University of California, Berkeley, CA, 94720-7450, USA}
\altaffiltext{3}{Sloan Research Fellow}

\begin{abstract}
We present a novel method for the optimal selection of quasars using time-series
observations in a single photometric bandpass.  Utilizing the damped random
walk model of \citet{kelly09}, we parameterize the ensemble quasar structure function in Sloan
Stripe 82 as a function of observed brightness. The ensemble model fit can then be evaluated rigorously
for and calibrated with individual light curves with no parameter fitting.  This
yields a classification in two statistics --- one describing the fit confidence and
one describing the probability of a false alarm --- which can be tuned, {\it a priori}, to achieve high quasar detection fractions
(99\% completeness with default cuts), given an acceptable rate of false 
alarms.  We establish
the typical rate of false alarms due to known variable stars as $\lessim 3$\% (high purity).
Applying the classification, we increase the sample of potential quasars relative to those known in Stripe 82 by as much as 29\%, and by nearly a factor of two in the redshift range $2.5<z<3$, where selection by color is extremely inefficient.
This represents 1875 new quasars in a 290 deg$^2$ field.
The observed rates of both quasars and stars agree well with
the model predictions, with $>99$\% of quasars exhibiting the expected variability profile.
We discuss the utility of the method at high redshift and in the regime of noisy and sparse data.
Our time-series selection complements well independent selection based on quasar colors
and has strong potential for identifying high-redshift quasars for BAO
and other cosmology studies in the LSST era.
\end{abstract}


\maketitle

\section{Introduction}
\label{sec:intro}

Active Galactic Nucleii (AGN) --- and quasars (QSOs) in particular ---
continue to play a central role in modern astrophysics.
AGN emission is the hallmark of supermassive black hole growth.  Powerful
AGN outflows of photons and matter affect the Universe on small
(accretion disk) and large (galactic, extragalactic) size scales, and
there is now substantial evidence both from observations and
simulations (both semi-analytical and hydrodynamical) that a ``quasar
mode'' is a key part of massive galaxy formation and evolution. On an
ensemble basis, the observed number densities \citep[e.g.,][]{croom09},
large scale-structure \citep[e.g.,][]{shen07,shen09,ross09},
and use as a cosmological probes via the Lyman-$\alpha$
forest and Baryon Acoustic Oscillations \citep[BAO; e.g.,][]{mcdonald07}, demonstrate that luminous AGN
--- despite their rare nature --- provide elucidating
constraints on some of the grandest questions of our time.

To fully exploit the potential of active galaxies, astronomers must identify and characterize 
the physics of large, representative samples.  Recent surveys 
have identified quasars using spectroscopy \citep[e.g.,][]{schneider10},
color-selection \citep[e.g.,][]{richards09},
X-ray detections \citep[e.g.,][]{bauer04}, and optical variability
\citep[e.g.,][]{s10}, among other methods.  The quasar catalogs
generated by these surveys exhibit a range of different completeness
and efficiency characteristics.
This is partly due to differences in
telescope sensitivities for the different surveys, but also to
intrinsic differences in the physics that each survey samples.  Deep
X-ray surveys and optical variability surveys currently show the most
promise for identifying large numbers of AGN per square degree of sky
\citep{brandt05}.

Optical surveys most often utilize quasar photometric colors to separate quasars
from field stars.  Typical ``completeness'' fractions --- the fraction
of retained spectroscopically-confirmed quasars --- for the selection
of quasars based on color are $\gtrsim 70$\% \citep[e.g.,][]{richards01}, although
this can decrease dramatically, to $10-50$\%, in the redshift range ($2.5<z<3$) where the
quasar UV color excess ($u$-$g$ color) is similar to that of stars \citep{richards06}
or for highly extinguished quasars.  The ``purity'' of selection, or the fraction of false
alarms due to spectroscopically-confirmed stars, can be similarly low.  Optical variability provides an independent
selection criteria and is becoming an increasingly
important survey technique.  Quasar selection based on optical variability is a key component of current and
upcoming missions, such as the
Panoramic Survey Telescope \& Rapid Response System 
\citep[\hbox{PanSTARRS};][]{kais02} and the Large Synoptic Survey
Telescope \citep[LSST;][]{iv08,ab09}.
Indeed, with the recent endorsement by the Astro2010 Decadal Survey
of both ground and space-based wide-field synoptic surveys, exploring
discovery in the time domain is particularly timely.

Quasar fluxes observed in optical passbands \citep[e.g.,][]{ms63} meander in
time, non-periodically, with flux differences that tend to be larger on larger timescales
\citep[e.g.,][]{hook94}.
This has historically been characterized in studies of auto-correlation using the
so-called ``structure
function'' \citep[e.g.,][]{Simonetti85,Hughes92}, evaluating the average square
variations versus timescale for an ensemble of quasars.  It is generally assumed
that the combination of observations from individual sources,
each perhaps observed only a pair or a few times, will accurately describe the
intrinsic variability of a given quasar.
Data from the Sloan Digital Sky
Survey \citep[SDSS; e.g.,][]{aba09},
and high-time-cadence observations for Stripe 82\footnote{The equatorial Stripe 82 region
(20h 24m $<$ R.A. $<$ 04h 08m, $-$1.27 deg $<$
Dec $<$ $+$1.27 deg, $\sim$ 290 deg$^2$)
was repeatedly observed --- 58 imaging runs from 1998
September to 2004 December --- with 1--2
observations per week, each Fall.}
in particular, have allowed major advances
in quantifying and understanding the nature of this variability
\citep[e.g.,][]{van04,iv04}, particularly for individual objects \citep[e.g.,][]{mac10,sesar10}.
We explore in detail below
the connection between the ensemble variability and the variability of individual,
well-sampled quasars.

The correlated variability of quasars is unique compared to most other variable
objects (e.g., stars) which tend on long-timescales (e.g., long relative to a periodicity timescale)
to exhibit non-correlated variability.
Quasars tend to vary much less on monthly and shorter timescales as compared to yearly timescales, unlike most stars (excluding, 
e.g., long-period variables).
This distinction motivates the possibility of using
quasar time series modelling to classify quasar and differentiate them from other objects.
Such a classification, particularly if it can be done efficiently, with few data,
could have tremendous benefit for selecting quasars for spectroscopic followup and
use in cosmological studies \citep[e.g., of BAO;][]{mcdonald07}.
Progress towards these ends is now becoming possible thanks to the generative ``damped random walk'' quasar
light curve model --- a model capable of stochastically producing a quasar
like light curve, in this case with only 2 input parameters --- uncovered by \citet{kelly09}.  \citet{kk10} have shown that
this model accurately describes the light curves of 100 well-sampled quasars, and it can
be used to separate these from stars.  \citet{mac10} have shown that the model accurately
describes individual quasars in Stripe 82.  We show that it can be used to accurately describe
the ensemble variability as well as the individual variability.

Below, we discuss a novel method to
fit the structure function for Stripe 82 using the damped random walk model
(Section \ref{sec:redux}).  We then show (Section \ref{sec:var})
how the fit of this average quasar model can be rigorously evaluated for individual light curves
to separate quasars from stars with no parameter fitting.  We show that nearly all known quasars ($>99$\%; Section \ref{sec:var_apply}) 
in Stripe 82 show the telltale
variability signature, with a very high completeness ($\approx 99$\%) that can be estimated a-priori.
This offers a substantial improvement over ad-hoc methods \citep[e.g., 90\% completeness in][]{s10} for
applying structure function fits to individual light curves.
The fraction of stars which could
be confused as quasars is, likewise, very small ($<3$\%) and is in reasonable agreement with {\it a-priori}~estimates.

The method we outline
below is based on maximum likelihood principles; it is therefore theoretically optimal and applicable to
any field or survey.  The quasar probabilities returned by the method robustly take into account model
uncertainties and uncertainties (both statistical and systematic) in the data.  The method
can be applied in the limit of very few data points (2 or more), because there are no free parameters
to fit and can provide key additional leverage to aid color-based selection schemes for future surveys.
The methodology is easily adaptable to include observations in multiple photometric filters and to
avoid contamination from spurious data.

\section{Data Selection and Ensemble Quasar Variability}
\label{sec:redux}

The majority of work presented herein makes use of the $ugriz$ photometry from the \citet{sesar07}
variable source catalog.  That catalog contains 67,507 $g<20.5$ objects 
in Sloan Stripe 82.  These are selected based
on the presence of statistically significant temporal variability
($>0.05$ mag RMS and $\chi^2_{\nu}/\nu >3$ in $g$ and $r$).
Additional details regarding the survey selection criteria can be
found in \citet{sesar07}.
The variability selection retains $\gtrsim 90$\% of known (spectroscopic) quasars in the field.
Those spectroscopically identified quasars were almost all entirely targeted as a result of color selection
from the main SDSS survey (and not because of time-domain characteristics).

Our goal is to differentiate quasars from other field sources using variability metrics alone,
without regard to color or cross-correlation with surveys at other wavebands.
We begin by evaluating the ensemble variability of 6304 
spectroscopically-confirmed quasars in \citet{sesar07} to quantify the observed
range of magnitude change, $\Delta {\rm mag}$, as a function of
timescale $\tau_{ij} = t_i-t_j$ between measurements at times $t_i$ and $t_j$.

\begin{figure}
\hspace{-0.2in}
\center{\includegraphics[width=6.0in]{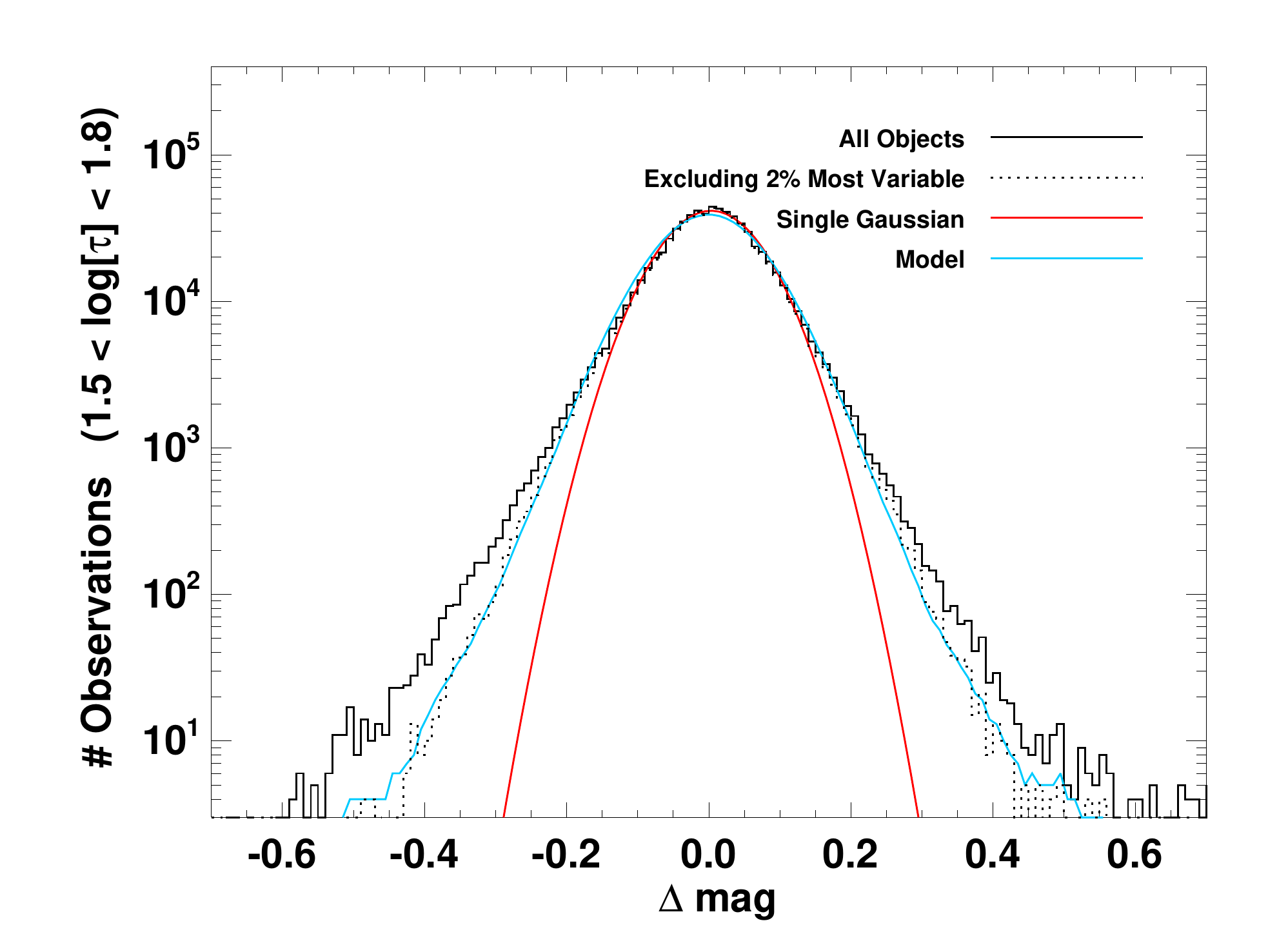}}
\caption{\small
The histogram of magnitude differences in $g$-band observed for 6304 
quasars on timescales $10^{1.5}<\tau<10^{1.8}$ days.  This is a time range which is particularly well-sampled for Stripe 82.  The distribution
has broader wings than a Gaussian (red curve); however it is 
well-represented by the superposition of Gaussian's (blue curve) with
widths  that
increase logarithmically (Table 1) with the quasar magnitude.
Residual non-Gaussianity can be attributed to a fraction 2\% of quasars
which exhibit excess variability.
}
\label{fig:sf_hist}
\end{figure}

\begin{figure}
\hspace{-0.2in}
\center{\includegraphics[width=5.0in]{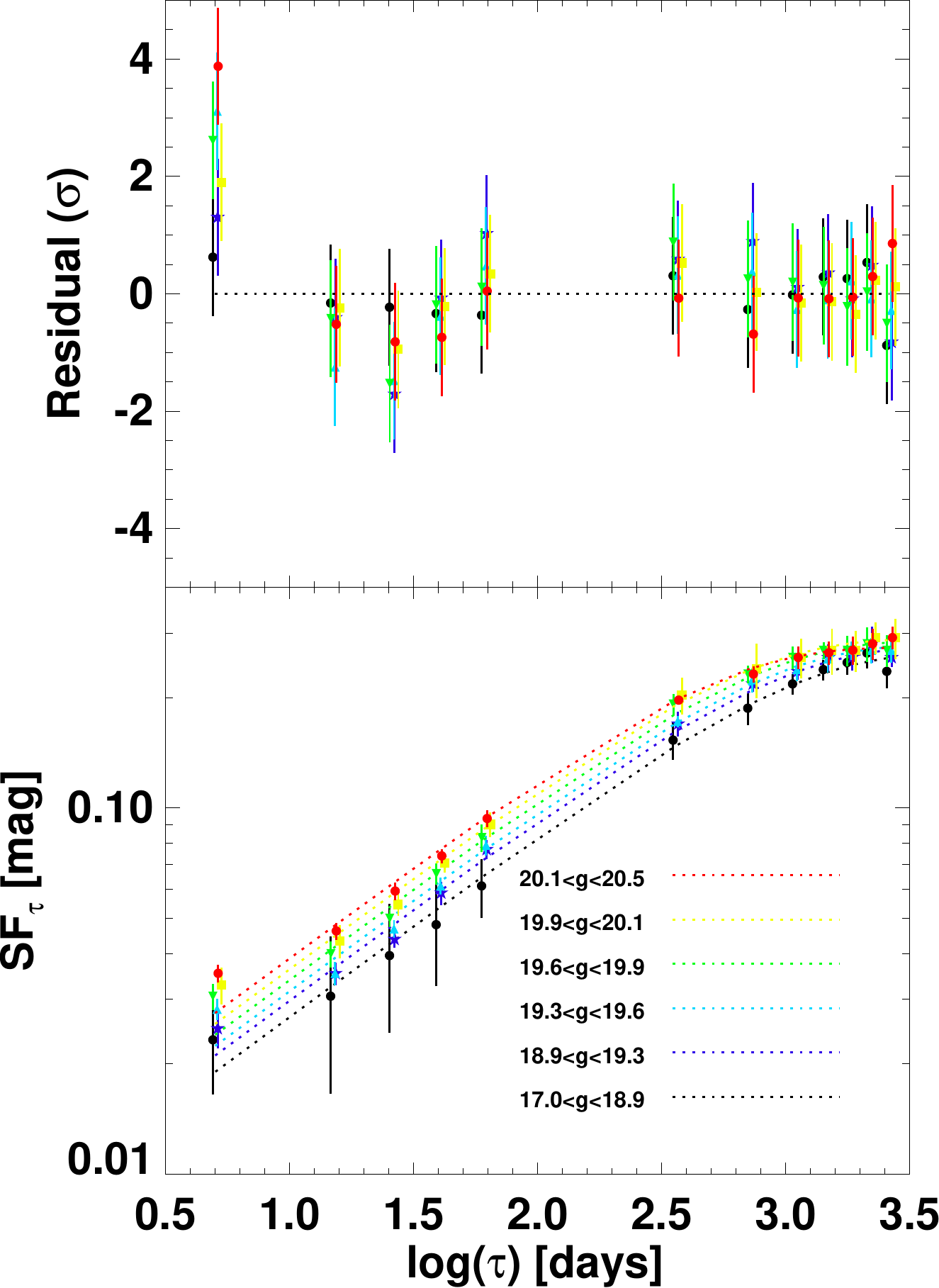}}
\caption{\small
The ensemble structure function $SF_{\tau}$ for quasars in SDSS Stripe 82 in $g$-band.
Each bin corresponds to a set of order $10^4$ difference measurements.  The
maximum likelihood Gaussian variance in addition to the measurement
uncertainty is determined for each pair, and the median and standard
deviation of these values for each set is plotted.
The data points are well-fit (dotted curves) by the damped random walk 
model
(see text) with parameters that vary with quasar magnitude only to
reproduce the observed increase in variability for faint sources
(Table 1).  There is modest evidence (shortest timescale
points) for untreated systematic measurement uncertainty at 
the $\lessim 1$\% level.}
\label{fig:sf}
\end{figure}

Figure \ref{fig:sf_hist} shows
a histogram of all magnitude differences in $g$-band for measurements
separated on timescales $\tau$ (in days) between $1.5<\log{(\tau)}<1.8$.  The
histogram is apparently symmetric, peaked at zero, and has broad, exponential wings.
The exponential distribution here is commonly found \citep[e.g.,][]{iv04}.
We find that the histogram can be well modelled as a sum of zero-mean
Gaussians whose widths scale logarithmically with the quasar magnitude.
Residual non-Gaussianity (i.e., an excess of large fluctuations as compared to the predictions from a Gaussian) is due to a small fraction (2\%) of sources
with excess variability (see also, Section \ref{sec:outlier}).
Figure \ref{fig:sf} plots the Gaussian width ---
after removing the measurement uncertainty ---  as a function of
magnitude and $\tau$.  This is commonly known as the first order quasar
structure function $SF_{\tau} = \sigma_{\Delta {\rm mag}(\tau)}$
\citep[e.g.,][]{Simonetti85,Hughes92}.

To move from the ensemble variability to the variability of a given
object, it is necessary to treat the covariances between neighboring
points --- which are estimated from the same data --- in the structure
function.  Given the approximate Gaussianity of the ensemble, it is
natural to fit a Gaussian random process to individual objects.
Consistent with previous works \citep[]{kelly09,kk10,mac10}, we find that the quasar
variability as a function of time difference is well-modelled using
a covariance matrix of the form:
\begin{equation}
C_{ij} = \sigma^2_i \delta_{ij} + {1 \over 2} \hat \sigma^2 \tau_{\circ} \exp{( -\tau_{ij}/\tau_{\circ} )},
\label{eq:omega}
\end{equation}
where $\sigma_i$ is the measurement uncertainty for the i'th observation, $\tau_{\circ}$ is an exponential damping
timescale (units of days), $\hat \sigma^2$ is the intrinsic variance between 
observations on short timescales $\tau_{ij} \approx 1$ day,
and $\delta_{ij}$ is the Kronecker delta function (1 for $i=j$, 0 otherwise).  This model
predicts $SF_{\tau} \propto \hat \sigma \tau_{\circ}^{1/2} [1-\exp{( -\tau_{ij}/\tau_{\circ} )})]^{1/2}$, which rises $\propto \tau_{ij}^{1/2}$ on 
short timescales ($\tau_{ij}\ll \tau_{\circ}$).  This
model is plotted over the data in Figure \ref{fig:sf}.  To
obtain an acceptable fit, we allow $\hat \sigma^2$ and 
$\tau_{\circ}$ to vary logarithmically with the median quasar magnitude.
The best fit scalings for the $ugriz$ bands are reported in Table 1.
Python software to calculate the fits and the quality statistics discussed below can be downloaded from the 
project webpage\footnote{http://astro.berkeley.edu/$\sim$nat/qso\_selection}.

\input{tab1.tex}

\begin{figure}
\hspace{-0.2in}
\center{\includegraphics[width=5.0in]{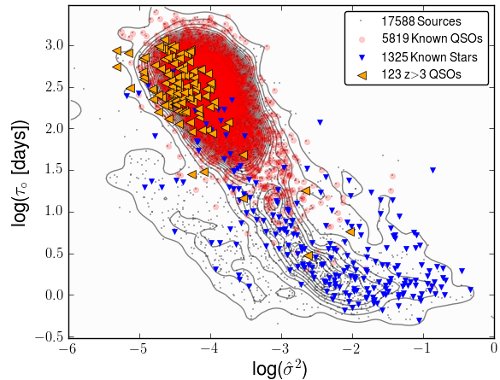}
\includegraphics[width=5.0in]{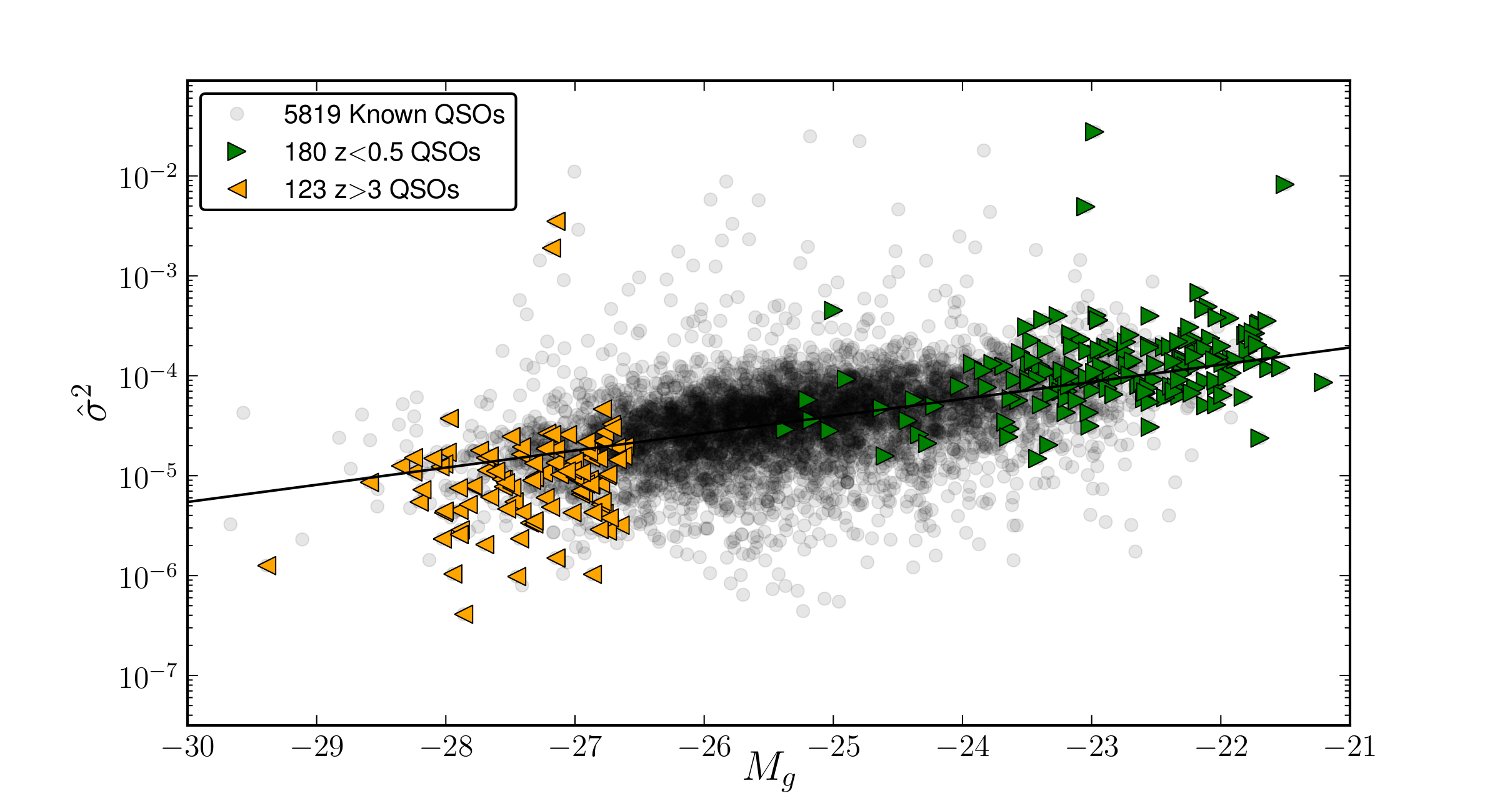}}
\caption{\small
(A) Exponential damping timescale $\tau_{\circ}$ for the random walk
$SF_{\tau}$ model versus the short-timescale variance $\hat \sigma^2$.  Spectroscopically-confirmed quasars
appear in the top left of the plot, while spectroscopically identified stars appear in the bottom
right.  High redshift ($z>3$) quasars tend to have weak, short-timescale variability (see text).  (B) The short-timescale variability scales with
intrinsic brightness $\log{(\hat \sigma^2)} = 0.17G-0.1$, with large scatter.  High redshift objects appear toward the left due to the survey flux limit $g<20.5$).  Low redshift objects appear toward the right.  Typical uncertainties are $\pm 1$ dex in $\tau_{\circ}$ and $\hat \sigma^2$
(90\% confidence).
}
\label{fig:var_tau}
\end{figure}

We can also fit the $SF_{\tau}$ model directly
by maximizing the posterior probability of the model given the data \citep[see, also,][]{kk10}.
The best-fit $\hat \sigma^2$ and $\tau_{\circ}$ values for each
quasar are plotted in Figure \ref{fig:var_tau}A for $g$ band.  Note that
consistent scalings with the $g$ band magnitude (Table 1) are found.

The origin of the scaling of variability with source brightness appears
to stem from the well-known anti-correlation between intrinsic brightness
and variability \citep[e.g.,][]{iv04}.  Figure \ref{fig:var_tau}B displays this trend
for the absolute magnitude in $g$-band, $M_g$.  We note that
$\hat \sigma^2$ no longer correlates with $g$-band magnitude or redshift after subtracting away this trend.
 High-redshift quasars
tend to appear toward the left of the plot (Figure \ref{fig:var_tau}B) 
--- weak short-timescale
variability --- due to the $g=20.5$ flux limit.  The effects of
a declining luminosity function (and potentially source evolution)
also work to keep low-redshift objects toward the right of the plot.
The normalization
of the fit in Figure \ref{fig:var_tau}B can be seen to vary slightly
with redshift also as a result of this flux limit, and we can expect
the apparent magnitude scalings (Table 1) to also have
some dependence on redshift and the survey selection (see, Section \ref{sec:redshift}).
These shifts are small compared to the (apparently intrinsic) $\sim$1 dex scatter in the scalings
(e.g., Figure \ref{fig:var_tau}B).

\section{Quasar Variability Selection Formalism}
\label{sec:var}

We can regard the parameterization of the $SF_{\tau}$ using the damped
random walk model as a rigorous mathematical approach to evaluating
the quasar likelihood for a given light curve, which takes into account 
all correlations in the data.  Employing the mathematical prescription
adapted from \citet{rp94} by \citet{kk10}, we write
for the probability of the data $x$ given the quasar variance
model $C = C(\hat \sigma^2,\tau_{\circ})$:
\begin{equation}
P(x|\hat \sigma^2,\tau_{\circ}) \propto
|C|^{-1/2} \exp{[-0.5 (x-x_{\circ})^T C^{-1} (x-x_{\circ})]}.
\label{eq:prob}
\end{equation}
We can marginalize analytically over $x_{\circ}$ and replace the
exponent in Equation \ref{eq:prob} with
\begin{equation}
-0.5\chi^2_{\rm QSO} = -0.5(x-x_{\circ,{\rm best}})^T C^{-1} (x-x_{\circ,{\rm best}}),
\end{equation}
where $x_{\circ,{\rm best}} = \sum_{i,j} C^{-1}_{ij} x_j / \sum_{i,j} C^{-1}_{ij}$.
The inverse of $C$ is tridiagonal \citep{rp94}, which allows for rapid $O(N)$
computation.
Given the parameterization above (Table 1) for 
$\tau_{\circ}$ and $\hat \sigma^2$ in terms of apparent magnitude mag,
we can then directly evaluate $\chi^2_{\rm QSO}({\rm mag})$ for all objects
of interest with no additional fitting (i.e., no fitting beyond the 
fitting of $x_{\circ}$).

For a quasar with the mean ensemble variability,
$\chi^2_{\rm QSO}$ should be $\chi^2_{\nu}$ distributed with $\nu$ degrees
of freedom, where $\nu$ equals the number of data points minus one.
The most likely value is $\chi^2_{\rm QSO}=\nu$.  The expected
 distribution of $\chi^2_{\rm QSO}$ for
a temporally-uncorrelated source can be evaluated by Monte Carlo or
estimated quickly as we now discuss.

\subsection{Significance Estimates}
\label{sec:signif}

The expected value for $\chi^2_{\rm QSO}$ for a source that is not a
quasar but
varies in a time-independent fashion with Gaussian scatter $\sigma_m$
is 
\begin{equation}
E[\chi^2_{\rm QSO}] \approx E[x^2] Tr(C^{-1})
 \equiv  \sigma_m^2 Tr(C^{-1}),
 \label{eq:null}
 \end{equation}
where $E[x]$ is the expectation value of $x$ and $Tr()$ is the
matrix trace operation.  In Equation \ref{eq:null}, we assume that all observations have approximately
the same uncertainty, and we ignore the light curve mean.  Because off-diagonal
terms in $C^{-1}$ do not contribute, $\chi^2_{\rm QSO}$ in
Equation \ref{eq:null}
is effectively a sum of the squares of $N$ Gaussian random
variables --- each with zero mean and standard deviation $\sigma_m$, which includes the
measurement uncertainty ---
multiplied by a constant, $Tr(C^{-1})/N$.  Therefore, the
quantity $N\chi^2_{\rm QSO} / [\sigma_m^2 Tr(C^{-1})]$
will be $\chi^2_{\nu}$ distributed, where the number of degrees
of freedom $\nu=N-1$.  (The missing degree of freedom represents
the light curve mean which we have neglected to write down).

The true value of $\sigma_m^2$ must be determined from the data.
In the case again of equal uncertainty on all data points,
\begin{equation}
P(\sigma_m^2|x) \propto \sigma_m^{-N} \exp{(-0.5 N v_x/\sigma_m^2)},
\label{eq:sigma_m}
\end{equation}
where $v_x \equiv \langle x^2 \rangle - \langle x \rangle^2$.
Multiplying the $\chi^2_{\nu}$ probability density for $N\chi^2_{\rm QSO} / [\sigma_m^2 Tr(C^{-1})]$ by
$P(\sigma_m^2|x)$ and integrating over $\sigma_m$, we find
a Beta distribution for the Null-hypothesis distribution of $\chi^2_{\rm QSO}$ given the data:
\begin{equation}
P(\chi^2_{\rm QSO}|x,{\rm not~quasar}) \propto (y[1-y])^{(\nu-1)/2},
\end{equation}
with $y \equiv \chi^2_{\rm QSO} / [\chi^2_{\rm QSO} + v_x Tr(C^{-1})]$.
This reference distribution approximates well the observed $\chi^2_{\rm QSO}$
frequencies for stars (Figure \ref{fig:qsofq_plot}) and can be used
as the reference distribution to calculate the significance of a given $\chi^2_{\rm QSO}$
value.  It is useful also to define 
\begin{equation}
\chi^2_{\rm False}/\nu \equiv {v_x Tr(C^{-1}) \over \chi^2_{\rm QSO}}
\end{equation}
The quantity $\chi^2_{\rm False}$ will be small and
of order $\nu$ for a time-independent, non-quasar variable source
(i.e., a potential false alarm).

\subsection{Confidence Estimates}

As we discuss above, the $\chi^2_{\rm QSO}$ calculated for
a hypothetical mean quasar --- varying according to
Equation \ref{eq:omega} with the best-fit parameters in Table 1
defining $\hat \sigma^2$ and $\tau_{\circ}$ --- will be of order $\nu=N-1$
and $\chi^2_{\rm QSO}$ will follow a $\chi^2_{\nu}$ distribution.
However, the variability of a given quasar may depart from 
the mean sample variability.  We can allow for this on a source-by-source basis
by replacing $C \rightarrow C/s^2$
in Equation \ref{eq:prob}.  The scale factor $s$ is a fudge factor that can be marginalized
over to allow the mean quasar model to acceptably fit each quasar light curve.

The distribution of $s$ given the data for a given quasar is given by Equation \ref{eq:sigma_m},
replacing $\sigma_m$ with $s$ and $N v_x$ with the most likely value for $\chi^2_{\rm QSO}$ of $\nu$.
In an argument parallel to that above used to derive significance, the $\chi^2_{\nu}$
probability density describing $\chi^2_{\rm QSO}/s^2$ can be convolved with $P(s^2|x)$
to find a reference Beta distribution to evaluate quasar confidence:
\begin{equation}
P(\chi^2_{\rm QSO}|x,{\rm quasar}) \propto (y'[1-y'])^{(\nu-1)/2},
\end{equation}
with $y' \equiv \nu / [\nu+\chi^2_{\rm QSO}]$.
Likewise for $P(\chi^2_{\rm QSO}|x,{\rm not~quasar})$ for stars,
this reference distribution approximates well the observed $\chi^2_{\rm QSO}$
frequencies for quasars (Figure \ref{fig:qsofq_plot}).

Note that because the reference distribution for evaluating significance has the
same form as that used to evaluate confidence, a curve of constant $\chi^2_{\rm QSO}/\chi^2_{\rm False}$
represents a curve of equal odds in favor of the hypothesis quasar versus not-quasar,
given equal prior information.

\section{Quasar Variability Selection Application}
\label{sec:var_apply}

In addition to 6573 quasars from \citet{schneider10}, 3020 labelled stars (i.e., spectroscopically-confirmed)
 in the
DR7 are plotted in Figure \ref{fig:var_tau}A.  We note that the total
number of objects now known to be quasars and used in these plots (and those below)
is larger than the quasar sample of 6307 from \citet{sesar07} used above
to establish $SF_{\tau}$. Known stars tend to have
$\tau_{\circ}\lessim 100$ days (i.e., approximately temporally-uncorrelated
variability) and also increased variability on short timescales
relative to quasars \citep[see, also,][]{mac10}.
In principle, Figure \ref{fig:var_tau}A can be
used for classification of unlabeled sources as well \citep[e.g.,][]{kk10}; however, we and others
find the uncertainties in these parameters to be large (often a factor of
ten even for well-observed sources).  Also, there is significant overlap
between the populations of labelled sources.  A more optimal separation
could be developed by seeking to fit no free parameters.

\begin{figure}
\hspace{-0.2in}
\center{\includegraphics[width=6.5in]{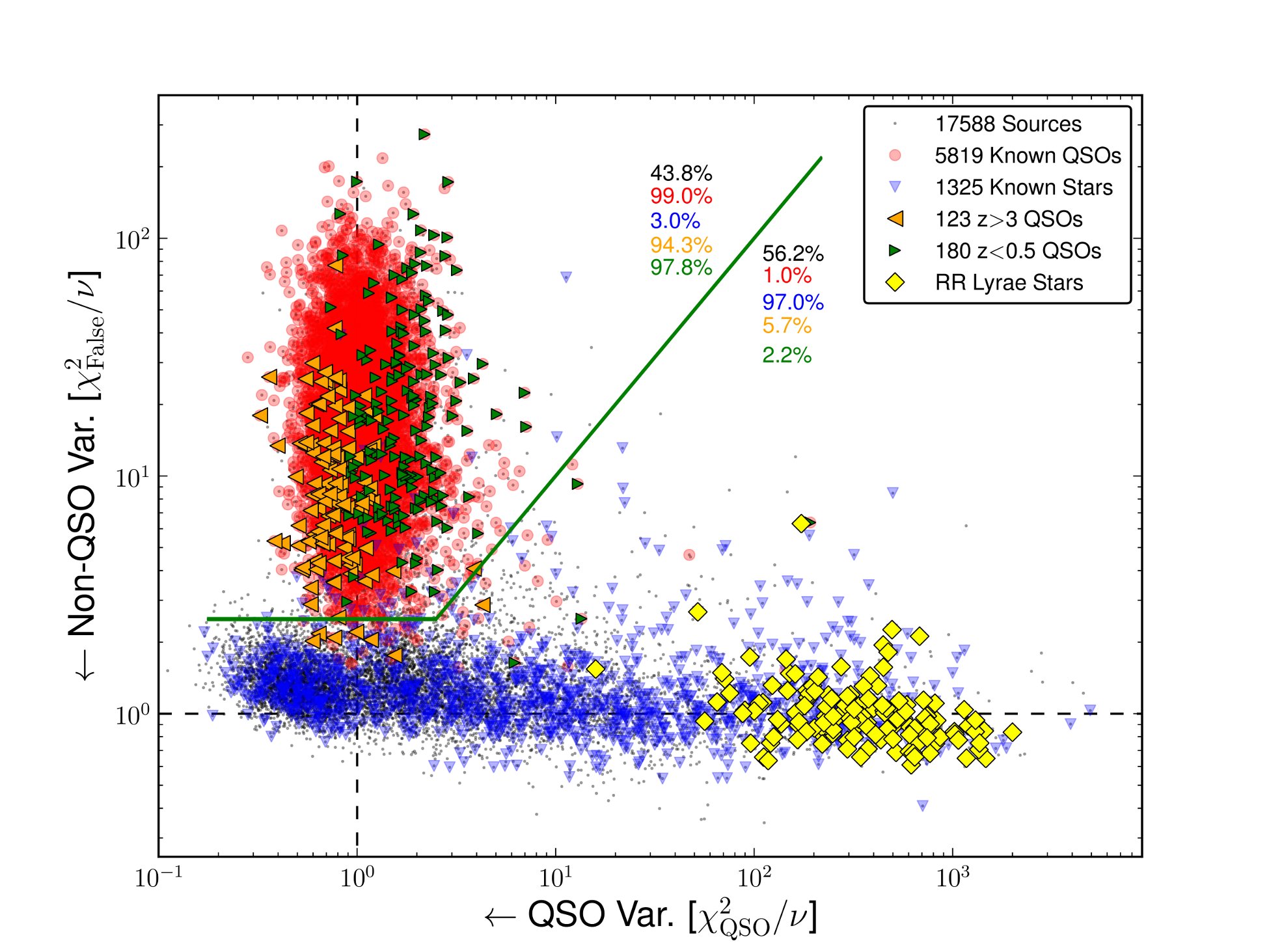}}
\caption{\small
Non-quasar variability metric $\chi^2_{\rm False}$ versus quasar-like
variability metric $\chi^2_{\rm QSO}$ for $g$-band observations
of sources with $>50$ epochs in \citet{sesar07}. 
Spectroscopically confirmed quasars \citep[from][]{schneider10} and stars are marked in red
and blue, respectively.  The low ($z<0.5$) and high ($z>3$) redshift quasars are additionally marked in green and orange, respectively, to display the trend toward low variability at high-$z$.
The solid line (highlighted in green) represents a cut 
based on equal probability
($\chi^2_{\rm QSO} = \chi^2_{\rm False}$)
between the hypotheses.   There is also a cut on $\chi^2_{\rm False}>2.5$,
corresponding to a $\gtrsim 3.5\sigma$ significance cut (Section \ref{sec:signif}) against false positives.
RR Lyrae stars \citep[from;][]{sesar10}, which are highly variable,
are also plotted as yellow diamonds.
The percentages of objects on either side of
the cut are recorded in the figure 
(descending in the same order as in the legend).}
\label{fig:qso_chi}
\end{figure}

Figure \ref{fig:qso_chi} shows $\chi^2_{\rm QSO}$ and
$\chi^2_{\rm False}$ for all 17,588 sources in \citet{sesar07} with 
50 or more observations in $g$-band.  Roughly 70\% of all sources
in Stripe 82 --- excluding 290 $<$ R.A.$<$ 340 degrees where the Galactic
stellar contribution dominates the source counts and blending between sources becomes common --- have 50 or more
observations.  There is a clear separation between
the majority of stars and quasars, and the values of $\chi^2_{\rm QSO}$
and $\chi^2_{\rm False}$ take on their expected values for labelled
sources.  For the population cut shown in Figure \ref{fig:qso_chi},
the completeness fraction for the fraction of known quasars retained
is 99\%.  One minus the fraction of known stars retained, the purity,
is 97\%.

Of all sources, we predict that 40\% are quasars, potentially increasing
the overall known quasar sample size by 20\%.  The candidate quasars
have a distribution in $g$-band magnitude consistent with that of the
known quasars.  We discuss the newly discovered quasars in more detail
below in Section \ref{sec:discuss}.

We also show explicitly in the plots the location of low redshift
($z<0.5$) and high
redshift ($z>3$) quasars, 97.8\% and 94.3\% of which, respectively, survive
the nominal quasar selection.  These tend to have systematically
high and low $\chi^2_{\rm QSO}$  values, respectively, due to the trend 
toward low variability at high luminosity discussed above.

The selection can be optimized to pursue $z>3$ quasars by placing a
more stringent cut on $\chi^2_{\rm QSO}\nu<2$ (low quasar-like variability)
to reduce the number of selected low-redshift quasars and also
relaxing the cut on overall variability to allow in more
false alarms, $\chi^2_{\rm False}/\nu>2$.  The resulting completeness (purity) for selecting $z>3$
quasars
is 97.6\% (95.7\%), considering only contamination by stars (95\% of $z<3$ quasars are retained, but most could be rejected using color information; Figure \ref{fig:color_plot2}).  Two of the three high-redshift quasars that do not survive the
refined $\chi^2_{\rm QSO}$ cut can be retained using
the outlier rejection scheme outlined below. (These quasars both have a single
deviant photometric point.)

\begin{figure}
\hspace{-0.2in}
\center{\includegraphics[width=6.0in]{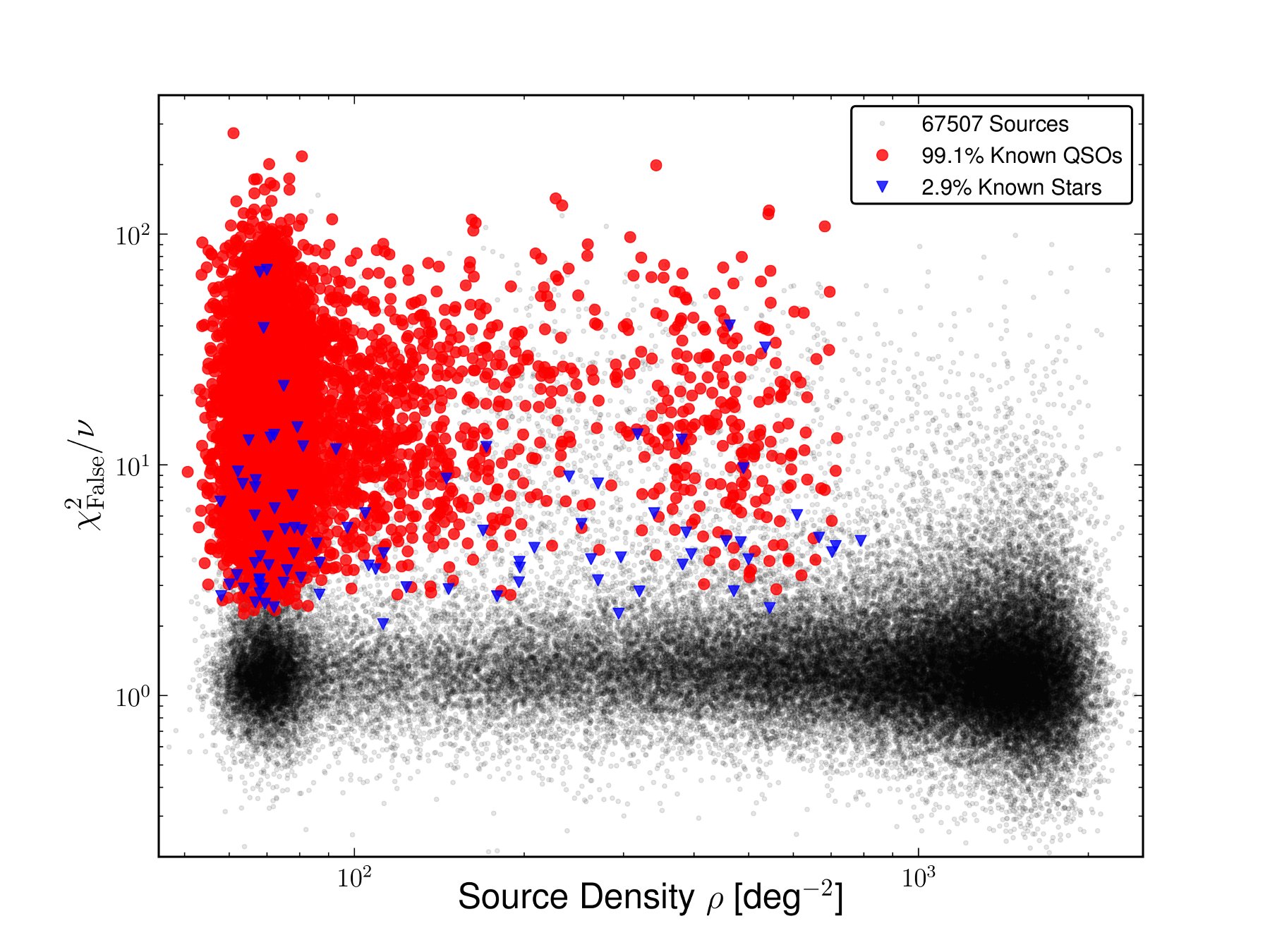}}
\caption{\small
The non-quasar variability metric versus source density $\rho$ for
all sources ($N\ge 9$ epochs per source).  Known quasars and
stars that survive a $>3.5\sigma$ significance cut on $\chi^2_{\rm False}$ 
and also have $\chi^2_{\rm QSO}<\chi^2_{\rm False}$
to be classified as quasars
are plotted in red and blue, respectively.
The classification --- which assumes equal prior probability
that a given source is or is not a quasar (Section \ref{sec:var_apply}) --- is quite stable over a broad
range in $\rho$.  At very high $\rho\gtrsim 10^3$ deg$^{-2}$ there are few
spectroscopic identifications and the stellar population begins to envelope
the quasar population, demanding a higher level significance cut.}
\label{fig:rho_plot}
\end{figure}

The classification defined by the cut in Figure \ref{fig:qso_chi} essentially
assumes equal prior probability that a source is or is not a quasar (Section \ref{sec:var_apply}).
We can see (Figure \ref{fig:rho_plot}) that the classification remains quite stable over
a broad range of source densities which include the high stellar-density region of 290 $<$ R.A.$<$ 340 (degrees) and also
now include sources with few measurement epochs (the minimum in the \citet{sesar07} sample is 9).
Because we are now classifying sources with few observation epochs, we select based on
the significance level of a given $\chi^2_{\rm False}$ rather than on simply $\chi^2_{\rm False}$.
The completeness (purity) we derive for Figure \ref{fig:rho_plot}
is 99.1\% (97.1\%), relatively little changed.  However, we caution that this neglects the
fact that few sources are spectroscopically identified
at high source densities $\rho\gtrsim 10^3$ deg$^{-2}$ 
(R.A. $\approx$ 300 degrees).  It is clear that the dramatic rise in the
overall source counts --- which must be dominated by stars --- would lead
to marked completeness and purity decreases.  It is, therefore, advisable
in such high source density regions to apply a more strict cut on
$\chi^2_{\rm False}$.
In principle, prior information on the expected
source density in a given direction for a given survey should be utilized
to dial in an appropriate threshold value.

\subsection{Significance/Confidence Validation: Tail Populations}

\begin{figure}
\hspace{-0.2in}
\center{\includegraphics[width=5.2in]{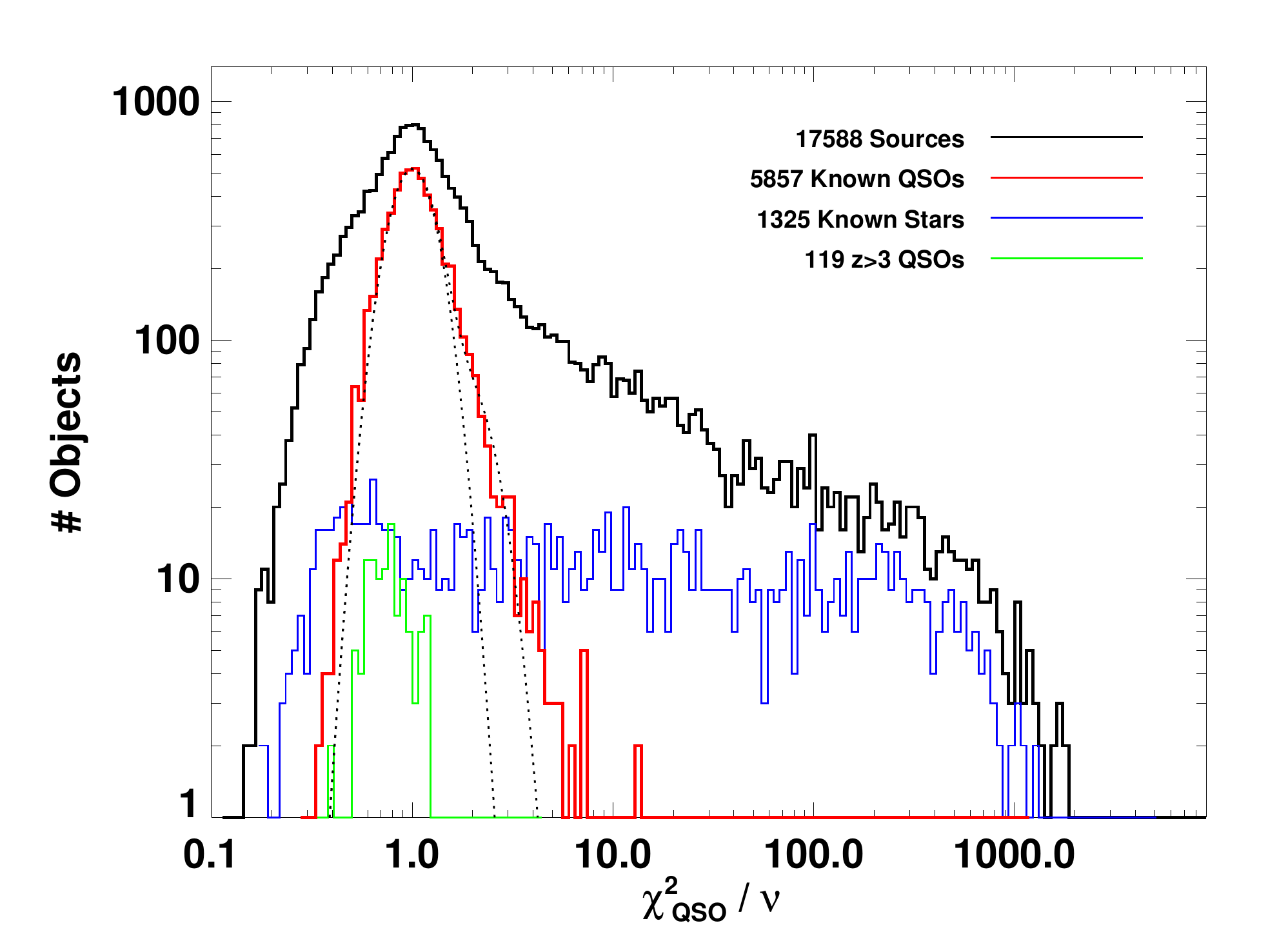}
\includegraphics[width=5.2in]{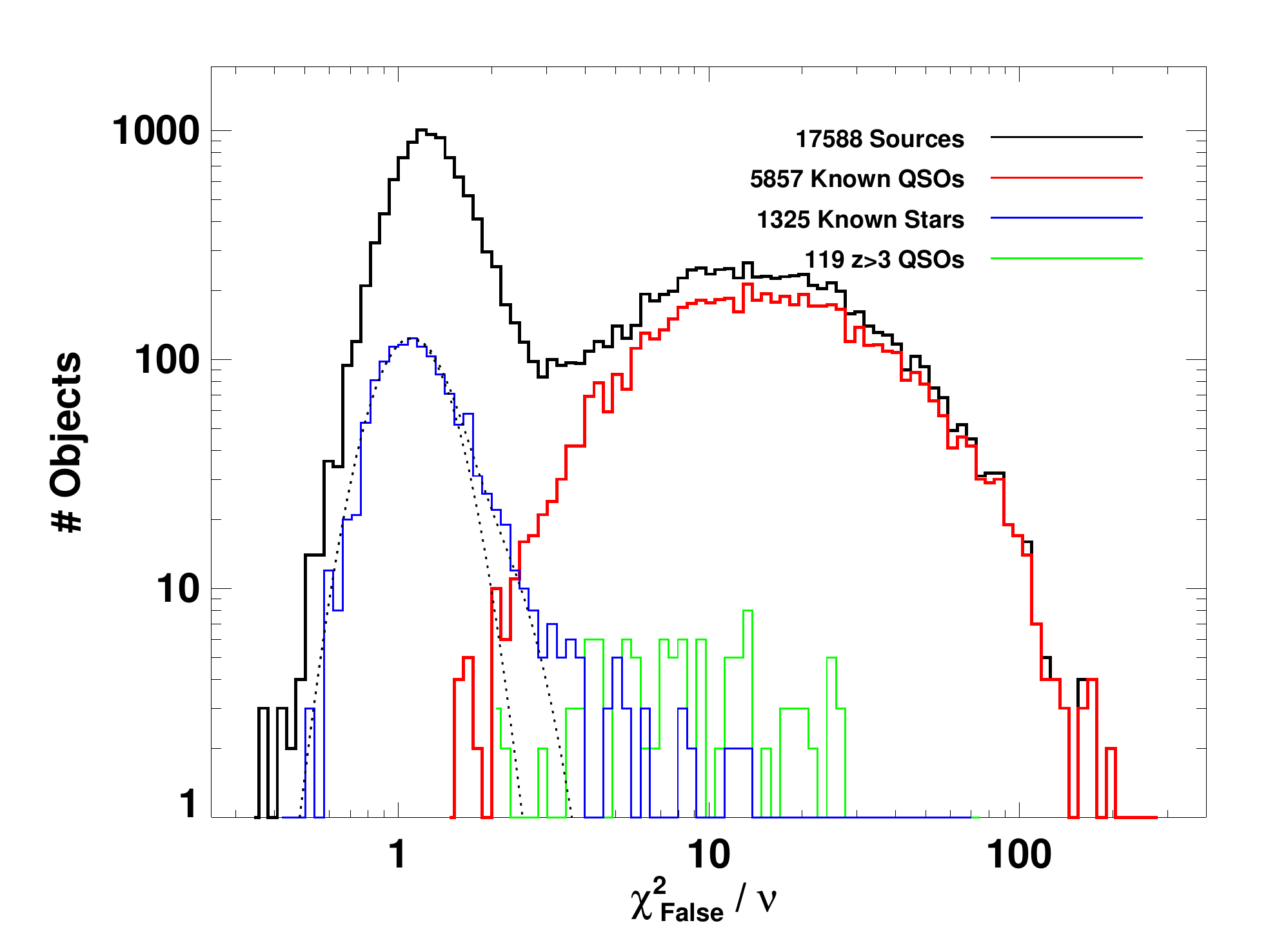}}
\caption{\small
Projections of the axes in Figure \ref{fig:qso_chi} for $\chi^2_{\rm QSO}/\nu$ (top) and $\chi^2_{\rm False}/\nu$ (bottom).  The dotted curves show the 
predicted Beta distributions for the source counts (see text), including
an excess of variable objects in both plots at the $\lessim 10$\% level.}
\label{fig:qsofq_plot}
\end{figure}

Figure \ref{fig:qsofq_plot} shows the observed distributions
of $\chi^2_{\rm QSO}$ and $\chi^2_{\rm False}$.  Over-plotted
are the expected Beta distributions (Section \ref{sec:var_apply})
for the median number of degrees
of freedom $\nu=55$.  These describe the observed frequencies
well, apart from a tail of 10\% of variable stars which have 
$\chi^2_{\rm False}/\nu \approx 2$ on average and a tail of 10\% of
highly-variable quasars which have $\chi^2_{\rm QSO}/\nu \approx 2$ on
average.  These tail contributions can be used to obtain a precise
significance estimate and rates which agree well with predictions,
although this improvement has little impact
on the logarithmic quasar/star separation in Figure \ref{fig:qso_chi}.
We note that 50\% of the $\chi^2_{\rm QSO}/\nu> 2$ objects are
at low redshift ($z<1$), where our approximation that variability
scales as apparent (instead of absolute) magnitude is expected to
break down.

Taking into account the excess number of events in the tails of the observed $\chi^2_{\rm QSO}$ and $\chi^2_{\rm False}$ distributions (Figures \ref{fig:qsofq_plot}),
a cut on either parameter at a value of 2.5 (corresponding to $\approx 3.5\sigma$ without the tail estimates) would need to be increased to $\approx 4.1$
to account for the tails at the same significance level.  This corresponds to a $\approx 5\sigma$ cut on the initial distributions, ignoring the tails.
Therefore, a conservative prescription to either reject false positives or to reject quasars, respectively, is to cut on $\chi^2_{\rm False}$ or $\chi^2_{\rm QSO}$,
respectively, at the $5\sigma$ significance level.  At this significance level, there are few outliers: 26 stars ($\approx 1$\% of the sample) masquerading with $\chi^2_{\rm QSO}<\chi^2_{\rm False}$ as
quasars and 15 ($\approx 0.3$\% of the sample) of overly-variable quasars, after excluding quasars at $z<1$.  We discuss the nature of these strong
outliers in Section \ref{sec:outlier} below.

\section{Discussion}
\label{sec:discuss}

\begin{figure}
\hspace{-0.2in}
\center{\includegraphics[width=6.5in]{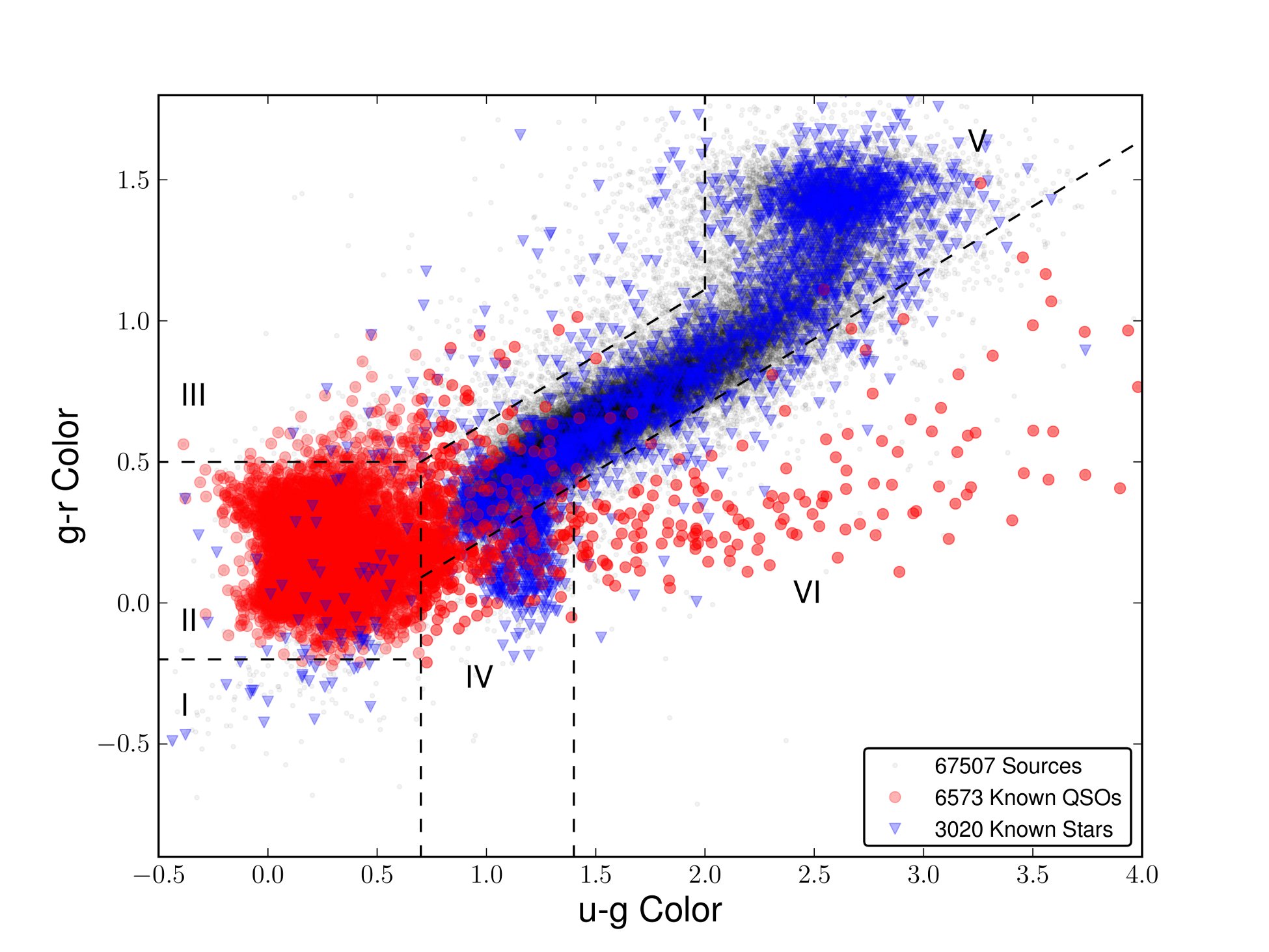}}
\caption{\small
Color-color plot showing the location of spectroscopically-confirmed stars (blue triangles) and quasars (red circles).  All of the sources from \citet{sesar07} are plotted as small black dots.  We demarcate with dashed lines the regions which \citet{sesar07} find to contain, primarily, white dwarves (I), low-redshift quasars (II), dM/WD pairs (III), RR Lyrae stars (IV), stellar locus stars (V), and high-redshift quasars (VI).}
\label{fig:color_plot}
\end{figure}

\begin{figure}
\hspace{-0.2in}
\center{\includegraphics[width=6.5in]{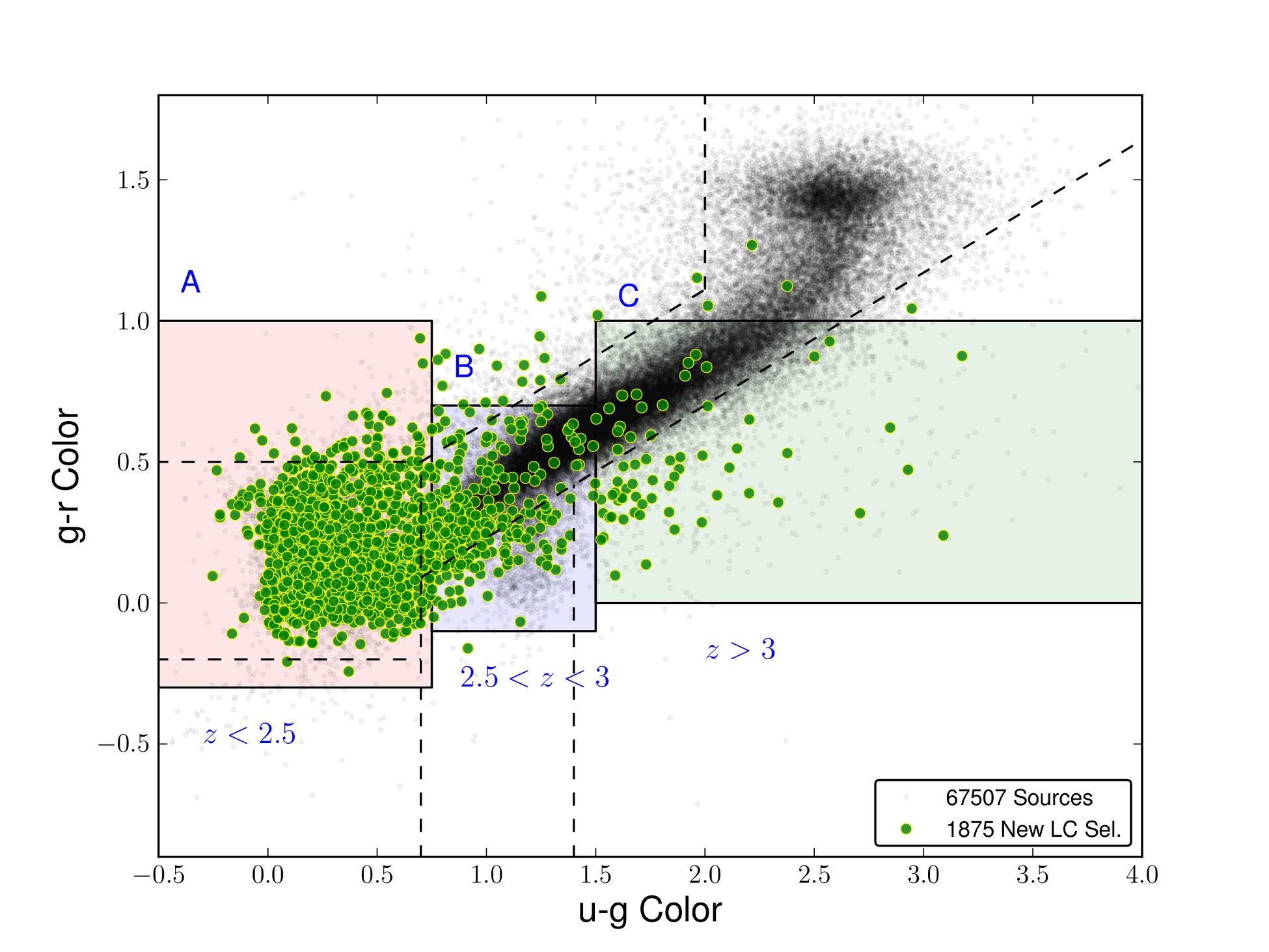}}
\caption{\small
Color-color plot showing the location of all sources from \citet{sesar07} in black (small dots).   We plot in green (circles) the location of 1875 new quasar candidates with significances $>5\sigma$, $>7\sigma$ for 290 $<$ R.A.$<$ 340 (degrees).  Labelled are the color-color regions A,B, and C, corresponding to quasars lying approximately in the redshift ranges $z<2.5$, $2.5<z<3$, and $z>3$, respectively.
Also plotted are the dashed-line demarcations from Figure \ref{fig:color_plot}.}
\label{fig:color_plot2}
\end{figure}

The light curve fitting above has the potential to provide vital 
information for quasar selection, particularly for those objects which
are challenging to select based on their photometric colors.
Figure \ref{fig:color_plot} shows the location of all \citet{sesar07}
objects in the $u-g$,$g-r$ color plane.  Low redshift quasars tend to lie
in a tight locus to the left of the plot.  Stars run along a branch
upward and to the left.  High redshift (and also potentially highly
extinguished) quasars fan out to the right of the quasar locus, through
the stellar region.  Color-based selection is clearly challenging for
these objects \citep[see also,][]{richards06}. We show in Figure \ref{fig:color_plot2} that
our method is capable of identifying a substantial number --- 1875 in this case ---
of highly statistically significant quasar candidates.
Very few of these ($\approx 1$\%) lie in the color-color space typically
dominated by stars (Region `V' in Figure \ref{fig:color_plot}), and 
we suspect some of these 26 candidates are extinguished quasars.

The selection for Figure \ref{fig:color_plot2} synthesizes 
recommendations from above.
For the high-Galactic latitude portion of Stripe 82 (excluding 
290 $<$ R.A.$<$ 340 degrees),
we apply a $5\sigma$ cut on $\chi^2_{\rm False}$ to eliminate the 
tail of variable stars
(Section \ref{sec:outlier}; Figure \ref{fig:qsofq_plot}, bottom).  
A higher significance
cut ($7\sigma$) is utilized for the low-Galactic latitude region to 
account for the higher stellar density (Figure \ref{fig:rho_plot}).  
Finally, we ignore sources
brighter than $i=18$ (corresponding to $\approx$3\% of the known 
quasar sample but
likely few of the candidate quasar sample).  This cut also helps to 
eliminate bright, red sources (i.e., primarily late-type stars 
at $g-r\gtrsim 1.3$, $u-g\gtrsim 2$ in Figure \ref{fig:color_plot}),
which are almost certainly not quasars.

\input{tab2.tex}

Quasars redshifts correlate strongly with their colors; three regions
(A, B, and C)
which contain $\gtrsim 90$\% of quasars in the redshift ranges $z\lessim 2.5$,
$2.5\lessim z\lessim 3$, and $z\gtrsim 3$, respectively, are plotted in 
Figure \ref{fig:color_plot2}.  These
regions are defined so as to maintain the total number of quasars
per redshift bin associated with a given color-color bin.
Table 2 displays the number of known,
spectroscopic quasars in each color bin as well as the number of
new candidate quasars discovered here.  
In the last two columns of the table, we make this comparison
separately for the bright ($i\le 19$ mag) and faint ($i>19$ mag)
samples.  In parentheses in these columns, we quote the implied fractional
increase in known quasars.

We potentially increase
the overall known quasar sample by 29\% (1875$/$6573), with a substantial
contribution ($+89$\%) in the color-color region (B) where color selection
is most difficult.
The number of known $z\gtrsim 3$ 
quasars would increase by 36--46\%, depending
upon whether we include 15 candidate quasars located in the stellar
region V in Figure \ref{fig:color_plot}.  These gains are
primarily due the selection here of the faint ($i>19$ mag) quasars,
although the fractional increases for $z\gtrsim 3$ quasars are
relatively independent of source brightness.  It is important to note when making these comparisons
that these gains are relative to {\it spectroscopically-confirmed}
quasars, whereas a large number of our candidates (1280) are also (un-observed) 
candidates based on color \citep{richards06}.  The fraction of new
quasars relative to color-selected quasars are quoted as
the second percentage in the final two columns of Table 2.  These numbers suggest that color-based selection is
complete at the $\gtrsim 95$\% level below $z\approx 2.5$ and rather
incomplete at higher redshift.

As most of the new quasar candidates come from the faint end of the detected source population, where the color selections are more incomplete 
(owing in part to the need for high signal-to-noise imaging in multiple filters), it is tempting to restrict to the brighter sources 
in making a direct comparison of time-domain and color selection completeness and purity; Table 2 provides this.
However, viewing the entirety of Stripe 82 as a fixed volume of data, the fact that high-confidence quasars can
be obtained fainter than the color-selection limit might be considered legitimate advantage of our technique.

\input{tab3.tex}

Table 3 gives the number of known and candidate
sources  corresponding to the 6 color-color regions discussed in
\citet{sesar07}.  These regions define the typical color-color locations of
various astrophysical transients (see, Figure \ref{fig:color_plot}).
The relative frequencies of the candidate quasars 
falling within a given region are roughly
consistent with those of the known quasars.  We note that there is a marked, relative increase
in the number of candidates possibly associated with stellar locus
stars (region V) and RR Lyrae stars (region IV).  RR Lyrae stars
tend to be strongly rejected as quasar candidates 
(Figure \ref{fig:qso_chi}).  In future work we will further explore what
fraction of
candidate quasars in the stellar locus are stars or highly
extinguished quasars.

\subsection{Weakly Variable or Non-Variable Quasars?}
\label{sec:weak}

\citet{sesar07} determine that 93\% of spectroscopically-confirmed quasars 
brighter than $g,r=19.5$ in Stripe
82 are variable at the $>0.03$ mag level.  They report a conservative fraction ($>90$\%),
which is limited by the measurement uncertainty.  Fitting a detailed model for the
expected variability, we can make a more general statement regarding the fraction of quasars that vary.
As reported above, $>99$\%
of quasars with 50 or more data points yield a highly significant $\chi^2_{\rm False}$.
The quasar population dwindles strongly below $\chi^2_{\rm False}/\nu=2.5$ (a $3.5\sigma$
cut on variability), unlike the stellar population.
The majority of stars have $\chi^2_{\rm False}\approx \nu$ which can occur either because
the variability is not qso-like or because the variability simply is not statistically
significant beyond the measurement error.

The fact that nearly all quasars brighter than $g=20.5$ have
$\chi^2_{\rm False}/\nu>2.5$ (99.1\% of quasars including those with as few as 9
measurement epochs; Section \ref{sec:var_apply}) indicates an intrinsic variability at least 60\% larger than
the typical measurement error of 0.02 mag.  Quasar with weaker variability can
be measured, and they are missing from Figure \ref{fig:qso_chi}.  The variability
cut in \citet{sesar07} is agnostic as to whether a source is truly a quasar; therefore
the presence of weakly variable stars where no weakly variable quasars are
found (in Figure \ref{fig:qso_chi}) indicates that approximately all quasars are
intrinsically variable.

Using the spectroscopic sample from \citet{schneider10}, in addition to the 6537 quasars considered above which have
light curves as part of the \citet{sesar07} project, there are 505 which are not contained in the \citet{sesar07} sample.
We downloaded\footnote{Using an SQL query on the Stripe 82 database hosted by SDSS at http://cas.sdss.org/stripe82/en/tools/search/x\_sql.asp}
the light curves for these quasars.
The extra fraction of potential quasars reflects uncertainty
in the overall sample size (due, e.g., to cuts on deblending flags, etc.).  
We find that 35 of 285 sources
with more than 8 measurement epochs yield $\chi^2_{\rm False}/\nu<2.5$.  Only 9 of these
have sufficiently weak variability to not be rejected as stable by a classical $\chi^2_{\nu}$
test at the $>5\sigma$ level.
This confirms that the
fraction of non-variable quasars is very small ($\lessim 1$\%) and
not impacted strongly by the \citet{sesar07} variability selection.

\subsection{Outliers}
\label{sec:outlier}

We have visually inspected the light curves and spectra for the 
strong outliers (26 stars and 15 quasars) identified above.  Two-thirds
of the quasar outliers appear to be the result of a mild amount
of errant photometry: we note the $\chi^2_{\rm QSO}\approx \nu$ 
when calculated in $r$ band for these objects and 
$\chi^2_{\rm QSO}\approx \nu$ also in $g$-band provided we remove
1--3 outlying (by $>5\sigma$) flux measurements.  This outlier 
identification
for individual points in a quasar light curve is determined using
the prediction for data point $x_i$ given all the other data points
\citep[see, e.g.,][or our software$^2$]{kk10}.  Of the remaining
five sources, two have poor photometry apparently as the result of
deblending from a nearby bright source and two have only marginally
large $\chi^2_{\rm QSO}$ after accounting for outliers.  We retain
one objects (SDSS J001130.40$+$005751.7), a ROSAT source which
appears to truly exhibit excessive, particularly short-timescale,
variability
($\chi^2_{\rm QSO}/\nu >30$, $\tau_{\circ}\approx 10$ days,  in this case)
as compared to the remaining sample of $\approx 6000$.

\citet{van04} find evidence for a significant increase
in short-timescale variability for ROSAT-associated quasars.
Consistently, we find a mild increase in $\chi^2_{\rm QSO}$ on
average for 94 ROSAT-associated quasars in \citet{schneider07}.
Of these, 80\% have $\chi^2_{\rm QSO}>\nu$, and the median
is $\chi^2_{\rm QSO}/\nu=1.5$.  All but a few of these are at $z<1$,
which suggests the poor quasar fit quality is due to the standard luminosity dependent variability
(Figure \ref{fig:var_tau}B) and not an anomalously high intrinsic variability.
\citet{mac10} have also looked at this issue and find no statistically significant
difference in variability for Stripe 82 quasars in the context
of the damped random walk model.

An effective scheme for rejecting outliers from an individual
quasar light curve, which retains the separation between quasars
and stars in Figure \ref{fig:qso_chi}, can be implemented as 
follows.  First and foremost, apply no outlier rejection in 
calculating $\chi^2_{\rm False}$.  Next, choose a maximum 
allowed number of outliers (say 3) and a significance level
for the residual (say $5\sigma$).  Calculate $\chi^2_{\rm QSO}$
and the model prediction for the light curve to evaluate outliers iteratively,
each time removing only the strongest outlier.  Once this is
complete, evaluate the standard deviation of the residuals
and repeat the process with a significance cutoff that is the larger
of 5 or 5 times the standard deviation.  This two-step
process makes it so that non-quasars receive little outlier rejection.

We note that the calculation of $\chi^2_{\rm QSO}$ with this outlier
rejection scheme substantially dampens by a factor $\approx 2$
the tail of 10\% of mild outlying quasars, leaving primarily only
the low-$z$ quasars/AGN.   The peak in $\chi^2_{\rm QSO}$ also shifts
down by about $10$\% for all quasars.
We note that we do not utilize this outlier rejection
in the paper because the $\chi^2_{\rm QSO}$ tails are adequately
small in our view.
Some form of outlier rejection may be quite important when using
less pristine photometry.

The outlier rejection method summarized above is tailored for quasars
and has no effect on the tail of 26 strongly outlying stars in
$\chi^2_{\rm False}$ or the 10\% tail of modestly outlying stars.
Of these objects, only 4 are late-type stars (which could, in principle
be rejected based on color as discussed above).  Seven of the objects
have quasar-like colors (Figure \ref{fig:color_plot}), have targeting
flags in SDSS associated with quasars, and have low-confidence ($<95$\% at best)
redshift determinations.  We regard these as potentially mis-labeled
stars.  Fifteen high-confidence stars remain,
and nine (about 0.3\% of the full sample) exhibit quasar-like
($\chi^2_{\rm QSO}/\nu<2$) variability.  We note that one of these objects
was targeted as a white dwarf (SDSS J234601.89$-$004255.5),
and in a future study we will probe in more detail the nature of the
other outliers and the potential that the underlying physics (e.g., the presence of an accretion disk) may
be similar to that of quasars.

The data from multiple filters can also be combined to
obtain more robust values for $\chi^2_{\rm QSO}$ and $\chi^2_{\rm False}$.
The average of these for $g$ and $r$ bands leads to a small, but significant, improvement in
the separation in Figure \ref{fig:qso_chi}. (The addition of other bands
helps little.)  To eliminate outliers in the case of approximately simultaneous data,
it may also effective to throw out observation epochs exhibiting strong color
variations, for example $>5$ times the standard deviation away from the median color $g-r$.
We observe that the quasar colors are relatively stable in neighboring photometric bands
as a function of time, although there is a mild dependence on redshift.
There are $\approx 30$\% color variations that occur --- presumably due the difference
in continuum and line variability --- for redshifts that place
the $\lambda = 2800 \AA$ Mg$_{\rm II}$ line \citep[also,][]{iv04}
in one of the passbands.

We have explored also simultaneously fitting Equation \ref{eq:prob} to the data
from 2 or more passbands to potentially improve statistics.  One way to do this is to
use the covariance matrix (Equation \ref{eq:omega}) for the data from one
reference passband (e.g., $g$-band) and then to allow the other bands to only
vary linearly with the $g$-band magnitude.  We find that the $g$ and $r$ band data can be fit-well in this fashion,
which further indicates the approximate constancy of quasar $g-r$ colors.
However, the statistics (as measured by the scatter and source separation in Figure \ref{fig:qso_chi})
do not improve.

\subsection{Redshift Estimation}
\label{sec:redshift}

In Figures \ref{fig:var_tau}B and \ref{fig:qso_chi} above, we show
a tendency for increased variability with decreasing luminosity which
maps to a trend of decreasing variability with redshift.  The
survey flux limit plays a role in this trend, and evolutionary effects
are also likely at work.  Our classification appears to perform best
above $z=1$ as a result.  In principle, the classification
can be further optimized to identify bursts at higher redshift
(see, e.g., Figure \ref{fig:qso_chi}) by performing a
strict cut on $\chi^2_{\rm QSO}$ (see, Section \ref{sec:var_apply}).

It is also possible to estimate
the redshift of a quasar using either 
$\hat \sigma^2$ or $\chi^2_{\rm QSO}$.
The scatter in $1+z$ estimated this way relative to the true $1+z$
is large, a factor of 2.  There are also likely strong selection
effects at play.  The flux limit prevents high redshift bursts from
exhibiting strong variability.  That is, the \citet{malm22} bias
induces a time-domain bias.
The survey color-based selection may 
also play a role.  We caution the reader that selection effects
that define a given survey may also strongly impact the utility of
this redshift estimation.  It is not clear at present that
useful redshift constraints can be obtained from our variability
measures.

\section{Conclusions}
\label{sec:concl}

We have explored a parameterization of the ensemble quasar variability structure function using the damped random walk model \citep{kelly09}.
This enables a statistically rigorous evaluation of the fit of an individual
quasar to the expected sample average variability profile.  The latter step provides, essentially, a classification
between objects undergoing quasar-like variability and objects exhibiting temporally uncorrelated variability.
Unlike previous work, the classification requires no parameter fitting, is essentially free from survey-specific peculiarities, and appears to be very robust in separating known variable stars
from quasars.  Nearly all ($>99$\%) known quasars show the expected variability profile and can be cleanly separated from
stars, with $\lessim $3\% contamination by rare variable stars.  

Our variability-selected quasar candidates span a range of redshifts,
including a factor of nearly two increase in rates for intermediate redshifts ($2.5<z<3$), where
color-based selection performs poorly \citep[e.g.,][]{richards06}.  The classification performs well also
at high-redshift, and can be tuned to yield $>98$\% completeness for $z>3$ quasars.
We potentially increase the sample of quasars in Stripe 82 by 10--25\%, depending upon whether we compare to
color-selected (but not spectroscopically-confirmed) quasars or to spectroscopically-confirmed quasars only.  Most
of the new quasars are faint ($i>19$ mag).  Relatively independent of these
considerations, we increase the quasar fraction by 33\% or more for $z\gtrsim 2.5$.
Software to perform the classification as well as the list of candidate quasars can be downloaded
from the project webpage$^2$.

This work has been conducted under the auspices of a broader project \citep[The Time-domain Classification Project;][]{bloom08},
a goal of which is to characterize the allowed range of variability versus timescale for the full diversity of astrophysical objects.
We have shown above how the temporal profile of quasars stands out against the signature of variable stars, predominantly
on long timescales but with strong power to separate strong variables from quasars on short timescale.  Future
work will apply the methods outlined above to select highly extinguished and high-redshift quasars for spectroscopic
followup and also to embed the methods discussed above withing the broader TCP framework of
variable object classification.  Of particular interest will be additional tests of schemes (outlined above) to reject photometric outliers and
to apply the classification in the limit of very few data points for wide-field, and in particular real-time surveys,
for example using data from the Palomar Transients Factory \citep{rau09}.

The ``one-versus-many'' classification framework shown here should prove to be an important pillar of any survey that relies on time-domain photometric observations for target selection (even if the targets of interest are not quasars).
It appears that with the appropriate selection of cadences, the time-domain identification of quasars is now more robust, complete and pure compared with color selections.
This has important implications for survey strategies of LSST, the Synoptic All-Sky Infra-Red survey \citep[SASIR;][]{bloom09}, and other time-domain projects (e.g., WFIRST): when faced with a choice between more colors and a wider field in a given filter, the latter option is preferred.  With deep imaging in one blue band ($U$ or $g$), coupled with synoptic imaging in a redder band, we believe that high-redshift quasars may be most effectively discovered.

\acknowledgments
NRB is supported through the Einstein/GLAST Fellowship Program (NASA Cooperative Agreement: NNG06DO90A).
JSB was supported by a grant from the National Science Foundation (``Real-time Classification of Massive Time-series Data Streams''; Award \#941742).

We acknowledge John Rice, Dovi Poznanski, Nic Ross, \u{Z}eljko Ivezi\'{c}, Chelsea MacLeod, Berian James, David Schlegel, Chris Kochanek, and Gordon Richards for useful conversations.  We also thank an anonymous referee for excellent comments and suggestions which improved the manuscript.
We thank B. Sesar et al.~for their excellent SDSS variable source catalog.
Funding for the SDSS and SDSS-II has been provided by the Alfred P. Sloan Foundation, the Participating Institutions, the National Science Foundation, the U.S. Department of Energy, the National Aeronautics and Space Administration, the Japanese Monbukagakusho, the Max Planck Society, and the Higher Education Funding Council for England. The SDSS Web Site is http://www.sdss.org/.
The SDSS is managed by the Astrophysical Research Consortium for the Participating Institutions. The Participating Institutions are the American Museum of Natural History, Astrophysical Institute Potsdam, University of Basel, University of Cambridge, Case Western Reserve University, University of Chicago, Drexel University, Fermilab, the Institute for Advanced Study, the Japan Participation Group, Johns Hopkins University, the Joint Institute for Nuclear Astrophysics, the Kavli Institute for Particle Astrophysics and Cosmology, the Korean Scientist Group, the Chinese Academy of Sciences (LAMOST), Los Alamos National Laboratory, the Max-Planck-Institute for Astronomy (MPIA), the Max-Planck-Institute for Astrophysics (MPA), New Mexico State University, Ohio State University, University of Pittsburgh, University of Portsmouth, Princeton University, the United States Naval Observatory, and the University of Washington.

\end{document}

%% file: tab1.tex
\begin{table}
\begin{center}
\caption{Structure Function Parametrization}
\vspace{1mm}
\begin{tabular}{lcccc}\hline\hline
Filter & $a_1$ & $a_2$ & $a_3$ & $a_4$ \\\hline
$u$ &$-$3.90 & 0.12 & 2.73 & $-$0.02 \\
$g$ &$-$4.10 & 0.14 & 2.92 & $-$0.07 \\
$r$ &$-$4.34 & 0.20 & 3.12 & $-$0.15 \\
$i$ &$-$4.23 & 0.05 & 2.83 &  0.07 \\
$z$ &$-$4.44 & 0.13 & 3.06 & $-$0.07 \\
\hline
\end{tabular}
\end{center}
{\small Notes: $SF_{\tau} = \hat \sigma \tau_{\circ}^{1/2} [1-\exp{( -\tau/\tau_{        \circ} )})]^{1/2}$, with $\log{(\hat \sigma^2)} = a_1+a_2 ({\rm mag}-19)$, $\log{(\tau_{\circ})}
= a_3 + a_4 ({\rm mag}-19)$. Statistcal uncertainties in the above parameters are of order $10^{-4}$ and are negligble.  Magnitudes are in the AB system, uncorrected for Galactic extinction.}
\label{tab:fits}
\end{table}

%% file: tab2.tex
\begin{table}
\begin{center}
\caption{Redshift Counts from Candidate Quasars Colors}
\vspace{1mm}
\begin{tabular}{l|c|c|c|c}\hline\hline
Color Region &  \# Known & New (\% Known) & New $i\le19$ & New $i>19$ \\\hline
A ($z\lessim 2.5$) & 6086  &  1498, {\it 297}  (25\%, {\it 5\%}) & 93, {\it 44} (6\%, {\it 3\%}) & 1405, {\it 253} (30\%, {\it 6\%})\\
B ($2.5 \lessim z\lessim 3$) & 325   &  288, {\it 230}  (89\%, {\it 71\%}) & 43, {\it 41} (58\%, {\it 55\%}) & 245, {\it 189} (98\%, {\it 75\%}) \\
C ($z\gtrsim 3$) & 140  &   65, {\it 48} (46\%, {\it 33\%}) & 13, {\it 11} (43\%,37\%) & 52, {\it 37} (47\%, {\it 34\%}) \\\hline
\end{tabular}
\end{center}
{\small Notes: The color regions A, B, and C are defined in the text and in Figure \ref{fig:color_plot}.
Columns 3--5 report the number of new quasars
relative to spectroscopically-confirmed quasars, followed after a comma by the number (italicized) of new quasars not already
color-selected in \citet{richards09} or spectroscopically-confirmed.}
\label{tab:candidate_colors}
\end{table}

%% file: tab3.tex
\begin{table}
\begin{center}
\caption{Source Region Counts from Candidate Quasars Colors}
\vspace{1mm}
\begin{tabular}{l|c|c|c|c}\hline\hline
Color Region &  {\footnotesize \# Known} & New (\% Known) & New $i\le19$ & New $i>19$ \\\hline
{\footnotesize I (white dwarfs)}  & 5 & 2, {\it 1} (40\%, {\it 20\%}) & 0, {\it 0} & 2, {\it 1} (no previous) \\
{\footnotesize II (low-$z$ QSOs)} &  5866 & 1387, {\it 239} (24\%, {\it 4\%}) & 79, {\it 34} (6\%, {\it 2\%}) & 1308, {\it 205} (29\%, {\it 5\%})\\
{\footnotesize III (dm/WD)} & 134  & 77, {\it 47} (57\%, {\it 35\%}) & 25, {\it 19} (50\%, {\it 38\%}) & 52, {\it 28} (62\%, {\it 33\%})\\
{\footnotesize IV (RR Lyrae)}  & 158 & 98, {\it 62} (62\%, {\it 39\%}) & 6, {\it 5} (25\%, {\it 21\%}) & 92, {\it 57} (69\%, {\it 43\%})\\
{\footnotesize V (stars)}  & 252 & 257, {\it 210} (102\%, {\it 83\%}) & 49, {\it 48} (80\%, {\it 79\%}) & 208, {\it 162} (109\%, {\it 85\%})\\
{\footnotesize VI (high-$z$ QSOs)} & 158 & 54, {\it 36} (34\%, {\it 23\%}) & 6, {\it 4} (16\%, {\it 11\%}) & 48, {\it 32} (40\%, {\it 27\%}) \\\hline
\end{tabular}
\end{center}
{\small Notes: The color regions I, II, III, IV, V, and VI correspond to the typical location of objects in parentheses in the first column (see Figure \ref{fig:color_plot}).
Columns 3--5 report the number of new quasars
relative to spectroscopically-confirmed quasars, followed after a comma by the number (italicized) of new quasars not already
color-selected in \citet{richards09} or spectroscopically-confirmed.}
\label{tab:source_colors}
\end{table}